\documentclass[a4paper,aps,pra,superscriptaddress, twocolumn,preprintnumbers,amsmath,amssymb,nobalancelastpage,10pt,longbibliography]{revtex4-1}

\usepackage{amsmath}
\usepackage{verbatim}
\usepackage{graphicx}
\usepackage{color}
\usepackage{braket}
\usepackage{subfigure}
\newcommand{\angstrom}{\textup{\AA}}

\usepackage[colorlinks=true,citecolor=blue,linkcolor=blue,urlcolor=blue]{hyperref}

\begin{document}


\title{Nanoscale phase separation and pseudogap in the hole-doped cuprates \\ from fluctuating Cu-O-Cu bonds}

\author{Sergi Juli\`a-Farr\'e}
\affiliation{ICFO - Institut de Ciencies Fotoniques, The Barcelona Institute of Science and Technology, Av. Carl Friedrich Gauss 3, 08860 Castelldefels (Barcelona), Spain}

\author{Alexandre Dauphin}
\affiliation{ICFO - Institut de Ciencies Fotoniques, The Barcelona Institute of Science and Technology, Av. Carl Friedrich Gauss 3, 08860 Castelldefels (Barcelona), Spain}

\author{Ravindra W. Chhajlany}
\email[Corresponding author: ]{ravi@amu.edu.pl}
\affiliation{Faculty of Physics, Adam Mickiewicz University, Umultowska 85,  61-614 Pozna{\'n}, Poland}

\author{Piotr~T.~Grochowski}
\affiliation{Center for Theoretical Physics, Polish Academy of Sciences, Aleja Lotnik\'ow 32/46, 02-668 Warsaw, Poland}

\author{Simon Wall}
\affiliation{ICFO - Institut de Ciencies Fotoniques, The Barcelona Institute of Science and Technology, Av. Carl Friedrich Gauss 3, 08860 Castelldefels (Barcelona), Spain}

\author{Maciej Lewenstein}
\affiliation{ICFO - Institut de Ciencies Fotoniques, The Barcelona Institute of Science and Technology, Av. Carl Friedrich Gauss 3, 08860 Castelldefels (Barcelona), Spain}
\affiliation{ICREA, Pg. Lluis Companys 23, 08010 Barcelona, Spain}

\author{Przemys{\l}aw R. Grzybowski}
\email[Corresponding author: ]{grzyb@amu.edu.pl} 
\affiliation{ICFO - Institut de Ciencies Fotoniques, The Barcelona Institute of Science and Technology, Av. Carl Friedrich Gauss 3, 08860 Castelldefels (Barcelona), Spain}
\affiliation{Faculty of Physics, Adam Mickiewicz University, Umultowska 85,  61-614 Pozna{\'n}, Poland}

\begin{abstract}
The pseudogap phenomenology is one of the enigmas of the physics of high-$T\rm _c$ superconductors. Many members of the cuprate family have now been experimentally characterized with high resolution in both real and momentum space, which revealed highly anisotropic Fermi arcs and local domains which break rotational symmetry in the CuO$_2$ plane at the intraunit cell level. While most theoretical approaches to date have focused on the role of electronic correlations and doping-induced disorder to explain these features, we show that many features of the pseudogap phase can be reproduced by considering the interplay between electronic and nonlinear electron-phonon interactions within a model of fluctuating Cu-O-Cu bonds. Remarkably, we find that electronic segregation arises naturally without the need to explicitly include disorder. Our approach points not only to the key role played by the oxygen bond in the pseudogap phase, but opens different directions to explore how nonequilibrium lattice excitations can be used to control the properties of the pseudogap phase. 

\end{abstract}

\maketitle

\section{Introduction}

 The physics of high-$T\rm_c$ cuprate superconductors is  one of the great  challenges of contemporary many-body physics. Independently of material details, high-$T\rm _c$ superconductors support a very rich and complex phase diagram~\cite{Keimer2015,Lee2008}. While the Mott insulator and the basic phenomenology of $d$-wave superconductivity itself are reasonably well understood, the nature of the metallic phase from which  superconductivity emerges is a  mystery of the high-$T\rm _c$ landscape. In particular, the origin of the pseudogap metal (PG)~\cite{Timusk_1999,Norman2005,sachdev2018topological}---a phase  with  highly suppressed low energy excitations  that appears as the hole doping is increased beyond the Mott insulator phase, and also  above the superconducting dome up to a characteristic temperature $T^*$---is a widely debated topic. The pseudogap has two complementary intriguing features: anisotropic Fermi arcs in momentum resolved photoemission spectra~\cite{Marshall96,RevModPhys.75.473,Vishik_2018} instead of closed Fermi surfaces expected of metallic states, and real-space nanoscale $C_4$ (discrete rotational) symmetry-breaking domains often associated with a local charge modulation~\cite{Kohsaka2007,Kohsaka2012,Fujita2014}. 

The Fermi-surface properties of the PG phase~\cite{Tremblay2006,Civelli2005,Ferrero2009,Sordi2012,Gunnarsson2015,Maier2002,Gull2013,Wu2018} have been theoretically linked to various mechanisms: topological order and spin liquid physics~\cite{Lee06}, phase incoherent $d$-wave superconductivity~\cite{MicnasRev,Randeria92,Emery1995,Alexandrov1996,Franz98,Berg07}, and the breaking of various electronic symmetries not necessarily related to superconductivity~\cite{Varma06,Chakravarty01,Honerkamp07,Zaanen89,Emery8814,Sachdev03,Kim08,Chubukov98,Kivelson1998}. A number of electronic correlation-based approaches~\cite{Vojta2009,Fradkin10}  predict nematic $C_4$ symmetry-breaking real-space orderings, where the organization of such phases into nanoscale domains is usually considered to arise from glassiness, i.e., the disordering effect of impurities~\cite{Nie2013,Lee2016}. While the main route  to explain the high-$T\rm _c$ phenomenology and its associated PG has been undertaken via electronic correlations, several effects suggest that the coupling to the lattice modes should not be neglected. These include the anomalous isotope effect~\cite{Hafliger2006}, the universal oxygen vibration frequency shift in the superconducting phase~\cite{Reznik95,Hewitt04,Pintschovius05}, and more recently the identification of the inequivalence of oxygen electronic and vibrational states in the two lattice directions of the CuO plane in the PG phase~\cite{Hinkov597,Daou2010}. Furthermore, experiments which drive the Cu-O bond to large displacements with resonant femtosecond laser pulses have shown evidence that a light-induced superconducting phase can be achieved for temperatures up to $T^*$~\cite{Kaiser2014}.

A development in this direction has been made through the modeling of fluctuating Cu-O-Cu bonds~\cite{Newns2007,Nistor2011,Hsiao2015}: these works were able to reproduce the $d$-wave superconductivity and some characteristics of the PG without electronic correlation effects. Interestingly, the fluctuating bond model (FBM) predicts a \textit{uniform} smectic/nematic oxygen bond order with $C_2$ spatial symmetry. The mechanism for its disintegration into the experimentally observed nanoscale domains remains, however, unclear.

In this work, we revisit the FBM and show that (i) its uniform smectic PG phase is intrinsically unstable towards macroscopic charge separation, (ii) it is therefore necessary to include effects of Coulomb interactions and consider the PG phase resulting from the interplay of bond-phonon instabilities and electron correlations, (iii) this interplay leads to a nanoscale phase separated PG in real space with a local $C_4$ symmetry-breaking bond order and Fermi arcs in momentum space, and  (iv) the nanoscale separation in this scenario does not result from quenched disorder. However, as reported in experiments, the PG is enhanced (reduced) by adding magnetic (nonmagnetic) impurities to the system. 

\section{Description of the Fluctuating Bond model} 
The  FBM describes the interplay of the buckling of anharmonically oscillating Cu-O-Cu bonds and hopping of electrons via a non-linear electron-phonon coupling. The  Hamiltonian  $H_{\text{FBM}}=H_{\text{el}}+H_{\text{ph}}+H_{\text{el-ph}}$ consists of the bare electron and phonon Hamiltonians, and the electron-phonon interaction. The bare electron Hamiltonian reads
\begin{equation}
H_{\text{el}}=-t_{0}\sum_{\langle i,j\rangle,\sigma}c^\dagger_{i,\sigma} c_{j,\sigma}+t'\sum_{\langle\langle i,j\rangle\rangle,\sigma}c^\dagger_{i,\sigma} c_{j,\sigma}-\mu\sum_{j,\sigma}n_{j,\sigma},
\end{equation}
where $c_{j,\sigma}$ ($n_{j,\sigma}$) is the electron annihilation (occupation) operator of a spin-$\sigma$ electron in the $3d_{x^2-y^2}$ orbital centered on site $j$, and $t_0$ and $t'$ are the nearest- and next-nearest neighbor hopping amplitudes. The bare phonon Hamiltonian is written as the sum over the bond oscillators,
\begin{equation}
H_{\text{ph}}=\sum_b \frac{p^2_b}{2 M}+\frac{\chi_0}{2}u^2_b + \frac{w}{16}u^4_b,
\end{equation}
where $M$ is the O mass and $u_b$ is its displacement perpendicular to the Cu-O-Cu nearest-neighbour bond $b$. The oscillator potential has a double-well structure with $\chi_0<0$ and $w>0$. A strong quartic potential for the Cu-O bond has been recently observed in coherent phonon experiments in Yttrium Barium Copper Oxide~\cite{Ramos2019}. 
The electron-phonon interaction couples the anti-bonding electron orbital charge $Q_b=\frac{1}{2}\sum_\sigma(n_{i,\sigma}+n_{j,\sigma}-c^\dagger_{i,\sigma} c_{j,\sigma}-c^\dagger_{j,\sigma} c_{i,\sigma})$ nonlinearly to the displacement $u_{b}$,
\begin{equation}
H_{\text{el-ph}}=-\frac{\nu}{2}\sum_b u^2_{b}Q_{b}.
\label{Hbc}
\end{equation} 

In this work, we show that the effects due to the interplay of $H_{\text{FBM}}$ and Coulomb interactions, which we consider as  maximally screened, i.e., via an on-site term $U\sum_{i}n_{i,\uparrow} n_{i,\downarrow}$, are of defining importance. These interactions are distinct from the long-range interactions between charges in antibonding orbitals ${\propto}Q_b Q_{b'}$ at different bonds considered in earlier works on FBM~\cite{Nistor2011,Hsiao2015}. 

\section{Mean-field decoupling of the electron-phonon interaction}

The large dimension of the Hilbert space of the Hamiltonian $H_{\text{FBM}}$ makes it impossible to treat with exact numerical methods: In addition to the square lattice of fermions, the motion of each O atom represents an additional continuum quantum degree of freedom. Hence, one needs to perform a series of approximations in order to extract the physics of the model.

Due to the large difference in electron and O masses, the motion of the latter on each bond can be treated as an oscillation around the quartic potential minima, which allows for a mean-field (MF) decoupling. One defines the mean-field FBM as 
\begin{equation}
H_{\text{FBM}}=H_{\text{FBM}}^{\text{MF}}+\Delta H_{\text{FBM}},
\end{equation}
where $H_{\text{FBM}}^{\text{MF}}$ differs from $H_{\text{FBM}}$ in the electron-phonon interaction term 
\begin{equation}
\begin{split}
H_{\text{el-ph}}^{\text{MF}}=&-\frac{\nu}{4}\sum_{b,\sigma} \langle u^2_{b}\rangle (n_{i,\sigma}+n_{j,\sigma}-c^\dagger_{i,\sigma} c_{j,\sigma}-c^\dagger_{j,\sigma} c_{i,\sigma})\\
&-\frac{\nu}{4}\sum_{b,\sigma}  u^2_{b} \langle n_{i,\sigma}+n_{j,\sigma}-c^\dagger_{i,\sigma} c_{j,\sigma}-c^\dagger_{j,\sigma} c_{i,\sigma}\rangle\\
&+\frac{\nu}{4}\sum_{b,\sigma} \langle u^2_{b}\rangle \langle n_{i,\sigma}+n_{j,\sigma}-c^\dagger_{i,\sigma} c_{j,\sigma}-c^\dagger_{j,\sigma} c_{i,\sigma}\rangle.
\end{split}
\label{Hbc}
\end{equation} 
Notice that the MF Hamiltonian consists of a quadratic electron Hamiltonian with renormalized bond-dependent hopping amplitudes $t_b = t_0 -\nu\langle{u^2_b}\rangle /4$, and a set of isolated phonon oscillators with renormalized bond-dependent $\chi_b=\chi_0+\nu\langle Q_b\rangle/2$. 
This MF system can be solved by finding, self-consistently, the values $\langle{u^2_b}\rangle$ and $\langle Q_b\rangle$ minimizing the free energy of the hole system (see Appendixes \ref{app:bogo} and \ref{app:loop} for details).

Finally, in order to benchmark the accuracy of the MF decoupling of the electron-phonon term, we have exactly solved a simplified system of a four-site lattice and compared the results of the two approaches (see Table~\ref{table:MF} in Appendix~\ref{app:table_MF_ED} for a quantitative analysis). The results show that the MF energy is higher than the one obtained through exact diagonalization (ED), but close to it, and that the effective hopping $t_{b}$ is also similar in the two approaches.
\section{Instability of the FBM}\label{sec:instability}

\begin{figure}[t]
  \centering
  \includegraphics[width=0.9\linewidth]{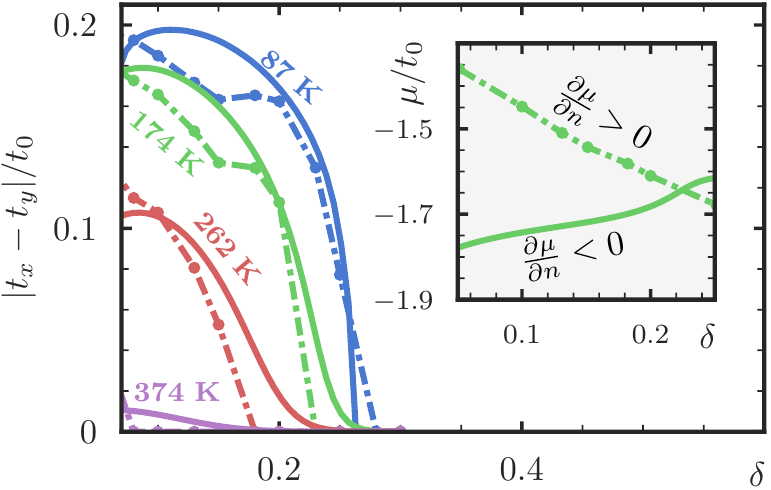}
\caption{Pseudogap phase as a function of hole doping and for different temperatures. The Figure shows the homogeneous MF parameter $\vert t_x-t_y\vert$ of the $H_{\text{FBM}}$ Hamiltonian on an $80\times80$-site lattice (solid lines) and the spatial average $|t_x-t_y|$ of the residual interactions model on a $30\times30$-site lattice (dashed lines). The inset depicts the chemical potential as a function of the hole doping at $174\ \text{K}$. We observe a negative compressibility $\partial\mu/\partial n<0$ in the homogeneous PG phase of $H_{\text{FBM}}$, which indicates the instability of this phase. On the contrary, the PG phase of $H_{\text{RI}}$ has a positive compressibility. The parameters of both Hamiltonians are fixed to $t_0=0.0083$, $t'=0.0011$, $\nu=0.03$, $w=0.17$, $\chi_0=-0.0025$, and $U=0$, where we use atomic units (energy  $E_0=27.2\  \text{eV}$ and length $a_0=0.53\ \angstrom$).
}\label{fig:restricted}
\end{figure}

The authors of Refs.~\cite{Newns2007,Nistor2011,Hsiao2015} found the spontaneous symmetry breaking $C_4$ to $C_2$ $\langle u^2_x \rangle \neq \langle u^2_y \rangle$ within a translationally invariant mean-field ansatz. From the electronic viewpoint, this is a bond ordered state with different hopping strengths $t_x \neq t_y$. The PG phase is then characterised by the splitting of the Van Hove singularity, which has an energy scale of the order of $\Omega_{\text{PG}} \propto |t_x - t_y|$. This leads to a strong reduction of the density of states between the Van Hove peaks. Figure~\ref{fig:restricted} shows the order parameter $\Omega_{\text{PG}}$ with respect to hole doping $\delta=1-n$ ($n$ is the electron density) at different temperatures (solid lines). We notice that the corresponding Fermi surface does not present Fermi arcs, which exhibit a $C_4$ symmetry.  Instead, the system only has a suppression of the spectral weight at $\mathbf{k}_{t_x>t_y}=(\pi,0)$ or $\mathbf{k}_{t_y>t_x}=(0,\pi)$. The authors of Ref.~\cite{Newns2007} suggested that impurities would form, in real space, domains of the two sectors of the symmetry breaking, leading to a restoration of the Fermi arcs.

\begin{figure}[t]
\includegraphics[width=0.95\columnwidth]{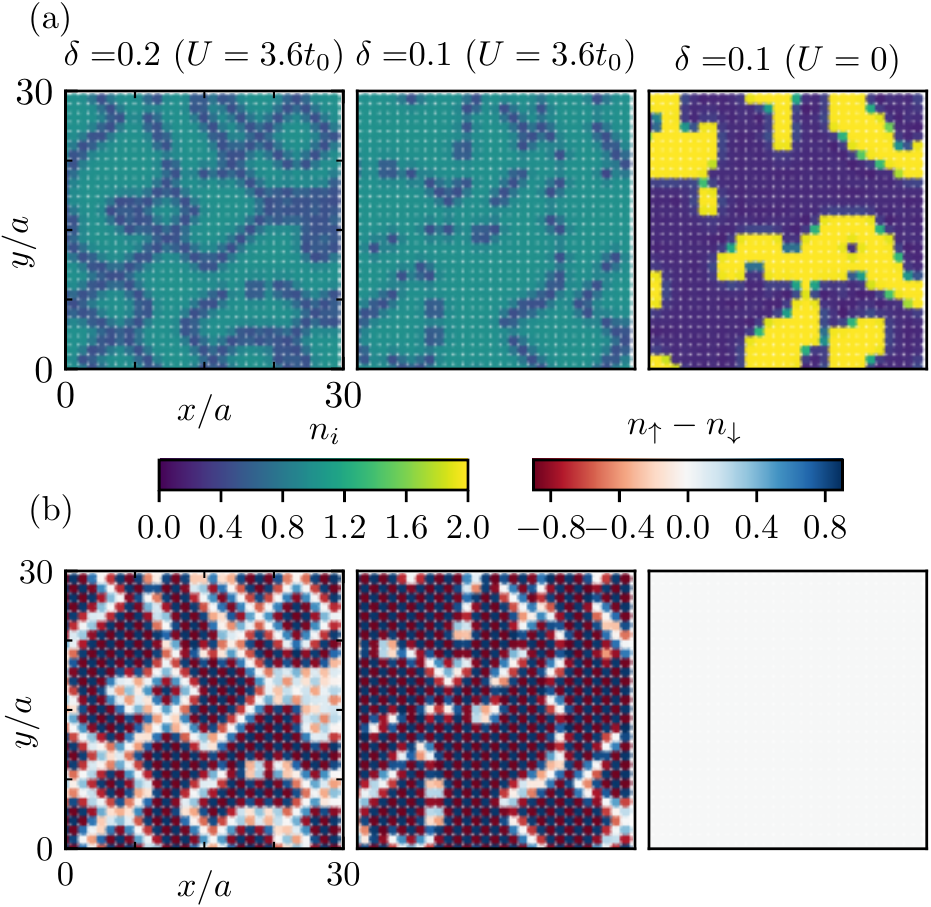}
\caption{Real-space features of $H_{\text{FBM}+U}$ at $K_BT=t_0/15$ for different dopings on a 30x30-site lattice. Parameters are set as in Fig.~\ref{fig:restricted}, except for the Hubbard, which is specified on each column. (a) Density plots showing microphase separation with small density amplitude (left and center) and macrophase separation with huge density amplitude (right). (b) Local spin polarization showing strongly polarized antiferromagnetic (AF) phase in the cases with finite $U$ (left and center).}
\label{fig:hartree}
\end{figure}

A more careful analysis nevertheless shows that this homogeneous PG solution is intrinsically unstable. The inset of Fig.~\ref{fig:restricted} shows that the compressibility \mbox{$\partial\mu/\partial n=-\partial \mu/\partial\delta$} is negative in the PG phase. We find this feature not to be specific to the  choice of FBM parameters but rather to persist for $\langle u^2_x \rangle \neq \langle u^2_y \rangle$ solutions. The effects of this instability can be visualized in real-space calculations using an unrestricted MF approach, in which the self-consistent averages $\langle{u^2_b}\rangle$ and $\langle Q_b\rangle$ are allowed to be independent for each bond.  
One then obtains macroscopic phase separation with distinct uniform regions of low and high electron density, without any bond order (see Fig.~\ref{fig:hartree} for $U=0$). 

\section{Inclusion of electron interactions: towards an effective model}

An important conclusion of the previous Sec.~\ref{sec:instability} is that Coulomb interactions are intrinsically needed to suppress the large charge imbalance of the FBM, and are therefore not only interesting from the point of view of competing phases (e.g., the charge density wave). A minimal extension of the FBM including Coulomb interactions leads to the Fermi-Hubbard model with bond phonons
\begin{equation}
H_{\text{FBM}+U}=H_{\text{e}}+H_{\text{ph}}+H_{\text{el-ph}}+U\sum_{i}n_{i, \uparrow}n_{i, \downarrow}.
\label{eq:fbm+U}
\end{equation}
A rigorous analysis of the FBM+$U$ Hamiltonian, for $U$ values typical for cuprate superconductors, constitutes a great challenge due to the strong electron correlations brought by the Hubbard term. In the following, we first discuss the numerical results obtained under different approximations. We then present an effective model that can be numerically studied in large clusters and leads to a stable pseudogap phase.

\subsection{Hartree-Fock study of the FBM+U}

We first study the effect of a large Hubbard repulsive $U$ on the phase separation with the unrestricted Hartree-Fock (HF) decoupling 
\begin{equation}
\begin{split}
&\left(n_{i,\uparrow} n_{i,\downarrow}\right)^{HF}= \braket{n_{i,\uparrow}} n_{i,\downarrow}+n_{i,\uparrow} \braket{n_{i,\downarrow}}-\braket{n_{i,\uparrow}} \braket{n_{i,\downarrow}}.
\end{split}
\label{Umf}
\end{equation} 
where we do not impose the translational invariance ansatz of Sec.~\ref{sec:instability}. The solution of the self-consistent equations (see Fig.~\ref{fig:hartree}) shows that, for a sufficiently large $U\gtrsim 3t_0$, the on-site interaction cures the macrophase separation generated by the electron-phonon interaction: the system exhibits smaller disconnected charge domains with lower density fluctuations. However, we do not observe any local $C_4$ symmetry breaking. This is due to the well-known overestimation of the magnetic correlations from the HF decoupling (see, e.g., Ref.~\cite{Fulde95}). In particular, the system has here a true gap with antiferromagnetic order at the relevant dopings and temperatures, as shown in Fig.~\ref{fig:hartree}(b), which masks any PG features.

\begin{figure}[t]
\subfigure{
\includegraphics[width=0.55\columnwidth]{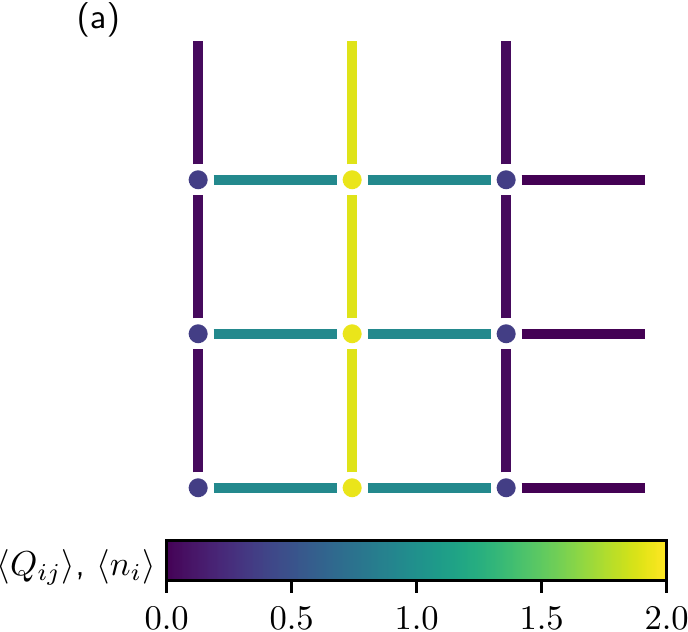}
}
\\
\subfigure{
\includegraphics[width=0.45\columnwidth]{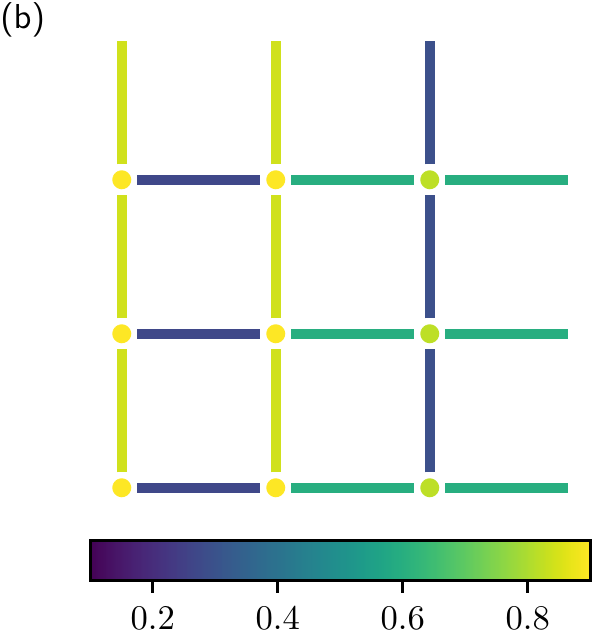}
}
\subfigure{
\includegraphics[width=0.45\columnwidth]{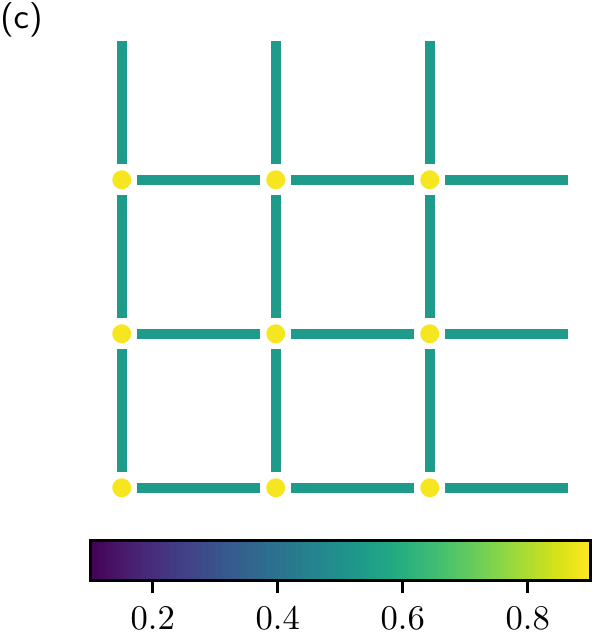}
}
\caption{Exact diagonalization results for $K_B T=t_0/15$. (a) $U=0$. (b) $U=3.6t_0$. (c) $U=3.6t_0,\ \nu=0$. We show the relevant local observables on a $3\times3$ lattice at electron filling $8/9$. The site colors encode the on-site occupation while the bond colors encode the values of the bond charge $Q_b$. The color scales are shown in the lower panel. }
\label{fig:ED}
\end{figure}
\subsection{Exact diagonalization study of the FBM+$U$}We now characterize more rigorously the PG close to half filling and in the presence of Hubbard interactions. To this end, we study the FBM+U model for a $3\times3$  cluster with periodic boundary conditions. We treat the  Hubbard interactions exactly and  the electron-phonon interactions with an unrestricted MF decoupling.  In Fig.~\ref{fig:ED}, for the unpolarized subspace of eight electrons (density $n=0.89$), we observe macrophase separation at $U=0$ with large density fluctuations through the lattice.  For a moderately large interaction $U=3.6t_0$, these fluctuations are strongly suppressed. Importantly, the C$_4$ symmetry breaking of the bonds is manifest and survives the formation of local magnetic moments.  

We emphasize that the exact treatment of the FBM+$U$ model for larger system sizes is numerically challenging due to the large values of $U$ typical of the cuprates. Nevertheless, we are here interested in the phonon bond order mechanism of the PG state and the associated generation of microphase separation, and the previous numerical results point to a scenario where  electronic correlations do not generate the PG phase but are essential to stabilize it.

\subsection{Residual interactions model}

We propose to discard the spatial fluctuations of the local density in the electron-phonon interaction as it would allow one to better treat larger systems without having the exaggerated effects of magnetic correlations at low hole doping. This effective model preserves the main effect of the repulsive interaction, which is to prevent macrophase separation.  One then obtains a model with at most a  residual small $U$ that now does not lead to magnetic order at temperatures relevant for the PG phase. We will see that this approximation reproduces qualitatively the ED results of the FBM+$U$ model, preventing the macrophase separation while allowing for a $C_4$ symmetry breaking.
 
 The resulting model, which we call the residual interactions (RI) model, differs from the FBM in the electron-phonon term, which is obtained by replacing the number operators $n_{i,\sigma}$ by the average density per spin species $\langle n_\sigma \rangle$ in the $Q_b$ of Eq.~\eqref{Hbc}. The latter gives rise to an effective 
 \begin{equation}
 \tilde{Q}_b =-\frac{1}{2}\sum_\sigma(c^\dagger_{j,\sigma} c_{j+1,\sigma}+\text{H.c.}),
 \end{equation}
 and a (total) density dependent renormalization of the quadratic part of the oscillator potential, 
 \begin{equation}
 \tilde{\chi}_0= \chi_0 - \nu/2 \langle n \rangle. 
 \end{equation}

\section{Pseudogap phase in the RI model}
In this section we analyze in depth the pseudogap phase of the RI model within MF+HF approximation with no translational invariance. 

\subsection{Fermi arcs and nanoscale domains}

\begin{figure}[t]
\centering
\includegraphics[width=1.0\linewidth]{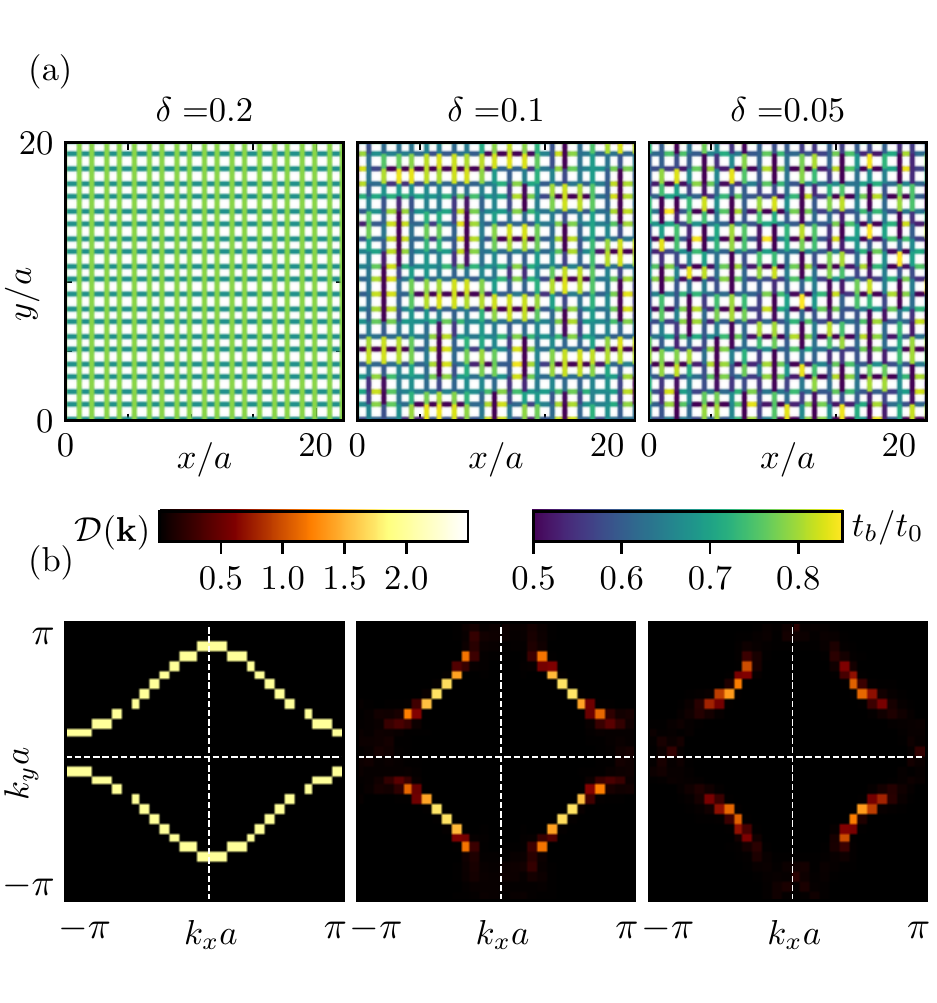}
\caption{PG dependence on hole doping for a fixed temperature $T=174\ \text{K}$ on a $30\times30$-site lattice for the RI model with the parameters of Fig.~\ref{fig:restricted}.  (a) Real-space plots of the effective electron hopping at each bond $t_b$. For $\delta=0.2$, the system presents an homogeneous $C_4$ symmetry breaking. For smaller dopings, we observe the formation of nanoscale domains with ladder structures. (b) Fermi surface $\mathcal{D}(\mathbf{k})$ in the Brillouin zone. We observe the appearance of Fermi arcs when increasing the hole doping.}
\label{fig:unrestrictedholes}
\end{figure}

\begin{figure}[t]
\centering
\includegraphics[width=1.0\linewidth]{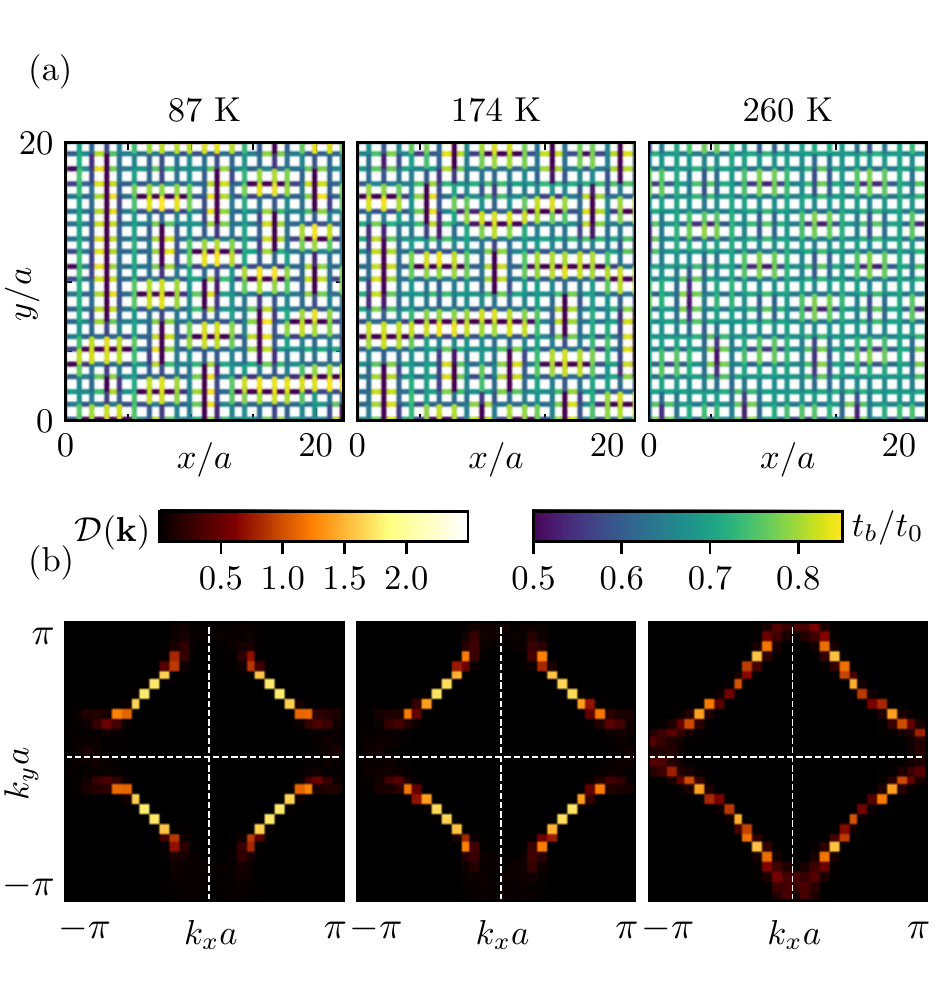}
\caption{PG dependence on temperature for a fixed hole doping of $\delta=0.1$ on a $30\times30$-site lattice for the RI model with the parameters of Fig.~\ref{fig:restricted} (a)  Real-space plots of the effective electron hopping at each bond $t_b$. The nanoscale domains are smeared out for increasing temperatures. (b) Fermi surface in the Brillouin zone. As the temperature is increased, the Fermi arcs evolve towards a closed metallic surface.}\label{fig:unrestrictedtemp}
\end{figure}

Figure~\ref{fig:restricted} shows the pseudogap parameters obtained for the unrestricted MF of the RI model for $U=0$ at different temperatures (dashed lines). These results are qualitatively similar to the ones obtained for the homogeneous solution of the FBM, but with a positive compressibility.  We now characterize more in depth the PG phase of the RI model. Figure~\ref{fig:unrestrictedholes} shows the PG dependence with respect to  hole doping for a fixed temperature. Figure~\ref{fig:unrestrictedholes}(a) shows the real-space distribution of the bond order parameter: 
for large doping, i.e., $\delta=0.2$, we observe a homogeneous $C_4$ symmetry breaking. Then, for smaller dopings, the system adopts a microphase separation with nanoscale domains, restoring on average the $C_4$ symmetry.
We also study the Fermi surface $\mathcal{D}(\mathbf{k})$ given by 
\begin{equation}
\mathcal{D}(\mathbf{k})=\sum_{j \in \mathcal{F}} |\braket{\mathbf{k}|\phi_j}|^2,
\end{equation}
where $\ket{\mathbf{k}}$ are the periodic Bloch states of the square lattice, $\ket{\phi_j}$ are the single-particle states of the unrestricted Hartree-Fock solution, and $\mathcal{F}$ is the subset of these states whose energy lies inside a window of width $t_0/10$ around the Fermi energy.
The results are shown in Fig.~\ref{fig:unrestrictedholes}(b). For $\delta=0.2$, close to the $C_4$ symmetry-breaking transition, the system is homogeneous and the Fermi surface is simply connected. The quasiparticle energies at the nodal points $\mathbf{k}=(\pm\pi/2,\pm\pi/2)$ are not affected by bond orderings, whereas at antinodal points, the dependence on bond orderings is stronger. Therefore, for small dopings where microphase separation occurs, the system presents nodal ``cold regions"~\cite{Ioffe}, forming characteristic anisotropic Fermi arcs, and strongly scattered ``hot regions" at anti-nodal points, resulting in a disconnected Fermi surface. This picture bears some similarity to the nematic glass theory~\cite{Kim08,Lee2016}, which however depends on external disorder.
The Fermi arcs' length increases with hole doping and leads to reconstruction of a simply connected Fermi surface close to the $C_4$-$C_2$ transition. The latter is in qualitative agreement with experimental observations~\cite{Keimer2015,Fujita612}. 

Figure~\ref{fig:unrestrictedtemp} depicts the dependence of the PG with respect to temperature for a fixed doping. For increasing temperature, a progressive closing of the Fermi arcs towards a metallic Fermi surface is observed.  In real space, the local amplitudes of the inhomogeneous $C_4$ symmetry breaking then become strongly suppressed.

\subsection{Role of impurities in the RI model}

The previous section shows that the nanoscale domains appear without the need of any type of quenched disorder. We now address the effect of non-doping impurities on the PG phase. These are often used as (destructive) probes of superconducting and PG properties of high-$T\rm _c$ materials. In particular, disorder is expected to destabilize nematic phases. However, two different behaviors are observed in experiments~\cite{Pimenov2005}: while substituting Cu for nonmagnetic Zn suppresses the PG,  substitution by magnetic Ni, remarkably, seems to have an enhancing effect on the PG energy scale. Here, we show that the results obtained from the RI model are in qualitative agreement with this impurity related phenomenology. 
\begin{figure}[t]
\subfigure{
\includegraphics[width=0.45\columnwidth]{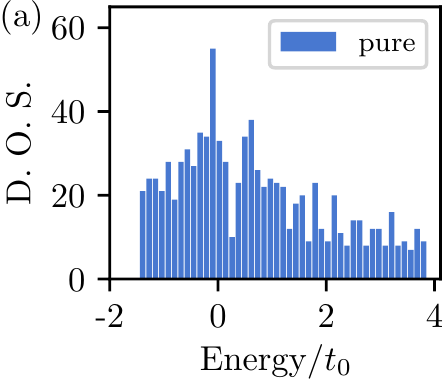}}
\subfigure{
\includegraphics[width=0.45\columnwidth]{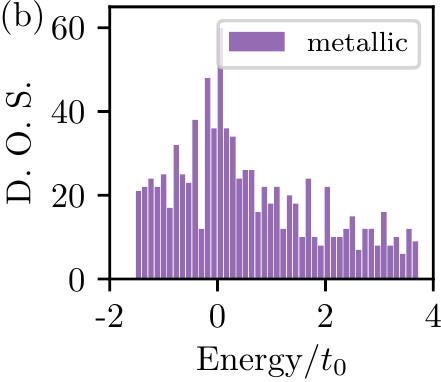}}
\\
\subfigure{
\includegraphics[width=0.45\columnwidth]{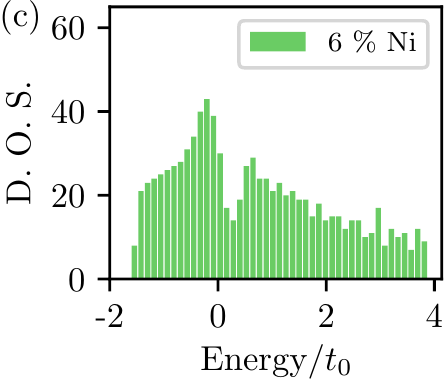}}
\subfigure{
\includegraphics[width=0.45\columnwidth]{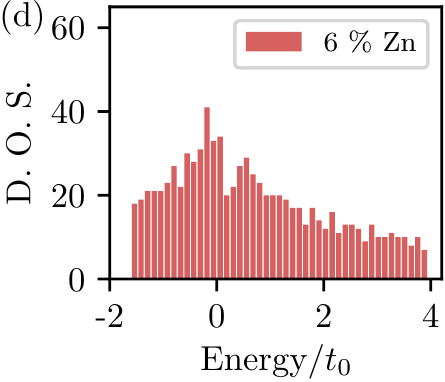}}
\caption{Density of states $\mathcal{N}(\omega)$ histograms for the cases considered in Fig. 4 of the main text. The number of bins is set to 45. For the residual interactions model (FBM+RI) we set the bare hole density $\delta=0.1$, $T=174\ \text{K}$, and $U/t_0=1.6$ on a $30\times 30$ lattice. The rest of the parameters are fixed to the same values of the main text.}\label{fig:dos}
\end{figure}
For Ni impurities, we use the Hamiltonian proposed by Va\v{s}\'atko and Munzar~\cite{Vasatko2016}
\begin{equation}
\begin{split}
H_{}=&-t_{0}\sum_{\langle i,j\rangle,\sigma}{\tilde c}^\dagger_{i,\sigma}{\tilde c}_{j,\sigma}+
J\sum_{\langle i,j\rangle}\left(\mathbf{S}_i\cdot\mathbf{S}_j-\frac{1}{4}n_i n_j\right)\\
&+E_{\text{Ni}}\sum_{\alpha}n_{\alpha}-4K\sum_{\alpha}\mathbf{S}'_{\alpha}\cdot\mathbf{S}_{\alpha},
\end{split}
\label{eq:Vasatko}
\end{equation}
where ${\tilde c}^\dagger_{i,\sigma}=c^\dagger_{i,\sigma}(1-n_{i,-\sigma})$ are the electron creation operators in the $3d_{x^2-y^2}$ orbitals projected such as to avoid double occupancy $n_i=\sum_{\sigma}c^\dagger_{i,\sigma}c_{i,\sigma}$, and $\mathbf{S}_i$ are the spin operators of the $d$ orbital. The Ni impurity sites are denoted as $\alpha$ and host additional $3d_{3z^2-r^2}$ orbitals. These orbitals carry a magnetic spin $\mathbf{S}_{\alpha}'$. The last term in Eq.~\eqref{eq:Vasatko} describes ferromagnetic Ni on-site interaction between $d$ orbitals. Considering an initial AF state polarized in the $z$ direction and in mean-field approximation, only the $S^z$ components survive, 
\begin{equation}
\mathbf{S}_{\alpha}\cdot\mathbf{S}'_{\alpha}\approx \langle S^z_{\alpha} \rangle S'^z_{\alpha}+S^z_{\alpha}\langle S'^z_{\alpha} \rangle-\langle S^z_{\alpha} \rangle \langle S'^z_{\alpha} \rangle. 
\end{equation} 
Since the $3d_{3z^2-r^2}$ orbitals are not affected by hopping, their spin within such approximation is classical. Nevertheless, the effect of these classical spins $\mathbf{S}'_{\alpha}$ cannot be considered as quenched disorder, as their equilibrium magnetization is determined self-consistently with the other spins $\mathbf{S_\alpha}$: at each step of the self-consistent loop, the requirement for $\langle S'^z_{\alpha} \rangle=1/2\ {\text{sgn}}(S^z_{\alpha})$ aligns it to the local $3d_{x^2-y^2}$ orbital magnetization $S_{\alpha}$ lowering the energy by 
\begin{equation}
-4K' S^z_{\alpha}\langle S'^z_{\alpha} \rangle\approx - 4K' (S^z_{\alpha})^2.
\end{equation} 

\begin{figure}[t]
\centering
\includegraphics[width=0.8\linewidth]{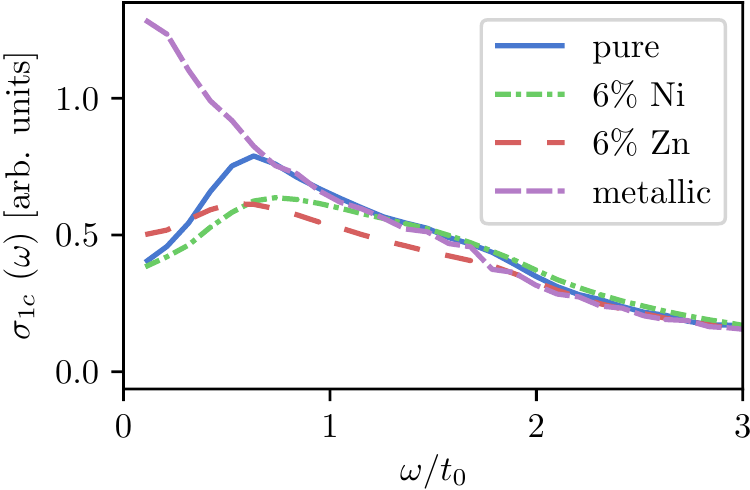}
\caption{Effect of impurities on the real part of the $c$-axis conductivity spectra in the residual interactions model for hole density $\delta=0.1$, $T=174\ \text{K}$, and $U/t_0=1.6$ on a $30\times 30$ lattice. The metallic solution (violet, dashed curve) is compared to the PG phase with and without impurities. The PG without impurities presents a characteristic peak. The latter is shifted to the left (right) for Zn (Ni) impurities.
 }\label{fig:caxis}
\end{figure}

The latter follows from that ${\text{sgn}}(4S^z_{\alpha})\approx 4S^z_{\alpha}$. As a consequence, we can consider the following effective Hamiltonian for Ni impurities: for the doped sites $\alpha$ we neglect the shift in the chemical potential proportional to $E_{\text{Ni}}$ and consider that the Hamiltonian is modified by the addition of the on-site term
\begin{equation}
\begin{split}
H_{\alpha,Ni}=&-4KS_{z, \alpha}^2=-K(n_{\alpha, \uparrow}-n_{\alpha, \downarrow})^2\\
=&-K(n_{\alpha, \uparrow}+n_{\alpha, \downarrow})  +2Kn_{\alpha, \uparrow} n_{\alpha, \downarrow},
\end{split}
\end{equation}
which leads to a modified on-site chemical potential 
$\mu_i\rightarrow \mu_i + K$ and Hubbard strength  $U_i\rightarrow U_i+2K$. $K$ is set to $3/4 t_0$. 

On the other hand, we denote the Zn-doped sites as $\lambda$, and we set $\mu_\lambda=\infty$ to effectively remove the doped site from the lattice \cite{Poilblanc1994}. In order to keep the hole concentration constant in the remaining available sites, we increase this quantity by $\tilde{\delta}=\delta+n_{\text{Zn}}$, $n_{\text{Zn}}$ being the concentration of Zn impurities.
 
 To quantify the effect of the above-mentioned impurities in the pseudogap unrestricted solutions we use the frequency-dependent transverse conductivity $\sigma_{1c}(\omega)$. The transverse conductivity is in general some combination of two parts, a momentum-conserving and a momentum-nonconserving part (see discussion in ~\cite{Prelovsek1998}). The hole doping in cuprates results in disorder in interlayer coupling since dopants can reside between the copper oxide layers. In this paper we focus therefore on the nonconserving part of the $c$-axis conductivity assuming that  interlayer tunnelings are in principle randomized both with and without Zn/Ni substitution, as in Ref.~\cite{Vasatko2016}. This $c$-axis conductivity contribution is  given by
\begin{equation}
\begin{split}
\sigma_{1c}(\omega) \sim \frac{1}{\omega}\int d\omega'\,&\left[ f(\omega'-\mu)-f(\omega'+\omega-\mu)\right]  \\ 
& \times\mathcal{N}(\omega')\mathcal{N}(\omega'+\omega)\, ,
\end{split}
\end{equation}
where $\mathcal{N}(\omega)$ is the density of states, and $f(\omega)$ is the Fermi-Dirac distribution. For completeness, we show in Fig. \ref{fig:dos} the density of states $\mathcal{N}(\omega)$ corresponding to the cases plotted in Fig. 4 of the main text.

 The $c$-axis conductivity results are shown in Fig.~\ref{fig:caxis} for both types of impurities, together for the pure case and a metallic solution, obtained as the self-consistent homogeneous mean-field solution with $C_4$ symmetry ($n_{i,\sigma} = n/2 \text{ and } t_b=t$).  The PG solutions show a characteristic low-energy suppression in the real $c$-axis conductivity spectrum as well as a peak. The PG energy scale $\Omega_{\rm PG}$ is often taken to be the peak position. It indeed behaves as advertised above. Furthermore, the depth of the suppression of the pure and Ni cases are similar, while the Zn PG is more filled in. \\

\section{Conclusions}

We have shown that including anharmonic Cu-O-Cu bond oscillations in Hubbard-type models leads to a number of key features of the PG phase including an inherent mechanism for nanoscale phase separation, Fermi arcs, and  appropriate response to defects. This points towards the fact that phonons play a key role in dictating the properties of high-$T\rm _c$ cuprates, and are not simply  secondary corrections to electronic correlation effects. Fundamentally, we therefore believe that our results will fuel  deeper investigations into the FBM+$U$ model, in particular via the treatment of electronic correlation effects more exactly beyond the mean-field approximation. Furthermore, it would be interesting to study the interplay of the electron-phonon interaction and the Coulomb interaction on the properties of the high-$T\rm _c$ superconductivity, within a non-transitionally invariant ansatz.  
Finally, the FBM+$U$ model could also serve as a natural basis to investigate how non-thermal and dynamical phonon distributions can be used to enhance and control phase competition in the cuprates. This would provide insights into the origins of light-induced non-equilibrium superconductivity and potentially lead to improved nonequilibrium control of the cuprates phase diagram. 

\section{Acknowledgements}

This work has been supported by the Spanish Ministry MINECO (National Plan 15 Grant: FISICATEAMO No. FIS2016-79508-P, SEVERO OCHOA No. SEV-2015-0522, FPI), European Social Fund, Fundacio Cellex, Generalitat de Catalunya (AGAUR Grant No. 2017 SGR 1341 and CERCA/Program), EU FEDER, ERC AdG OSYRIS and NOQIA, ERC StG SEESUPER, EU FETPRO QUIC, and the National Science Centre, Poland-Symfonia Grant No. 2016/20/W/ST4/00314. A.D. was financed by a Juan de la Cierva fellowship (IJCI-2017-33180). R.W.C. acknowledges funding from the Polish National Center via Miniatura-2 Program Grant No. 2018/02/X/ST3/01718.

\appendix

\section{Bogoliubov inequality}\label{app:bogo}

The mean-field treatment approach is based on the Bogoliubov inequality. We express the exact FBM Hamiltonian as $H_{\text{FBM}}=H_{\text{FBM}}^{\text{MF}}+\Delta H_{\text{FBM}}$. The Bogoliubov inequality reads
\begin{equation}
    F_{\text{FBM}} \leqslant F_{\text{FBM}}^{\text{MF}}+\langle \Delta H_{\text{FBM}}\rangle_{\text{MF}},
    \label{eq:bogo}
\end{equation}
where $F$ is the thermodynamic free energy, and the thermal ensemble of $H_{\text{FBM}}^{\text{MF}}$ with partition function $Z_{\text{FBM}}^{\text{MF}}$ is used to compute the expectation value $\langle \Delta H_{\text{FBM}}\rangle_{\text{MF}}$ and the free energy $F_{\text{FBM}}^{\text{MF}}=-K_BT\ln Z_{\text{FBM}}^{\text{MF}}$. The problem then reduces in finding the equilibrium state $\ket{\Psi_0}$ of $H_{FBM}^{MF}$ minimizing the right-hand side of Eq.~\eqref{eq:bogo}. Notice that $\ket{\Psi_0}$ will only contain MF correlations between electrons and phonons, and that it will satisfy the constraint $\bra{\Psi_0}H_{\text{FBM}}^{\text{MF}}\ket{\Psi_0}=\langle H_{\text{FBM}}^{\text{MF}}\rangle$. These conditions can be used to find $\ket{\Psi_0}$  within the self-consistency iterative algorithm described in Appendix~\ref{app:loop}.

\section{Self-consistent mean-field + Hartree-Fock loop}\label{app:loop}

Here we discuss the self-consistent loop used to determine the equilibrium state of $H_{\text{FBM}+U}$, equivalent to the FBM for $U=0$. After the MF+HF decoupling, the electron Hamiltonian has a quadratic form $\tilde{H}_{\text{el}}$ with renormalized bond-dependent hopping amplitudes $t_b = t_0 -\nu\langle{u^2_b}\rangle /4$ and on-site chemical potential $\mu_{i,\sigma}= -U\braket{n_{i,\bar{\sigma}}}+\frac{\nu}{4} \sum_{b\in i}\langle u_b^2 \rangle$. The phonon Hamiltonian $\tilde{H}_{\text{ph}}=\sum_b \tilde{H}_{\text{ph}}^b $ consists of 
a set of isolated phonon oscillators $b$ with renormalised bond-dependent $\chi_b=\chi_0+\nu\langle Q_b\rangle/2$. 

The MF parameters $\langle u^2_b\rangle$, $\langle Q_b\rangle$, and $\braket{n_{i,\sigma}}$ are found with a self-consistent iterative loop. Before starting the iterative algorithm, we have fitted the value of $\langle u_b^2 \rangle$ as a function of $\langle Q_b\rangle$ at a given temperature $T$. For this purpose, we have used a local phononic basis of $800$ states to find the eigenstates of $\tilde{H}_{\text{ph}}$ for $200$ values of $\langle Q_b\rangle$ in the interval $[0,2]$. These eigenstates are then used to compute the thermal expectation value of $\langle u_b^2 \rangle$ according to the Boltzmann distribution. Finally, a simple fitting routine is used to extract $\langle u_b^2 \rangle$ as a function of $\langle Q_b\rangle$ from the $200$ values obtained.

Once the fitting for the phonons has been performed, the iterative algorithm proceeds as follows: the initial conditions are imposed in the bond phonons, with an initial distribution for each variable $\langle u_b^2 \rangle$, and to the electronic density, with an initial density distribution $\braket{n_{i,\sigma}}$. At each iteration step, the single-particle states of $\tilde{H}_{\text{el}}$ are obtained, and from them the fermionic state at temperature $T$ and filling $n$ is constructed. From this fermionic state, one obtains the new distribution for $\braket{n_{i,\sigma}}$, and $\langle Q_b\rangle$, which gives the new value of $\langle u_b^2 \rangle$ through the previously fitted function. In order to avoid oscillating solutions, the update of the mean-field parameters is done progressively as
\begin{equation}
\langle \cdot \rangle_{i+1}=(1-\eta)\langle \cdot \rangle_{i} + \eta\langle \cdot \rangle_{i}^{\text{new}}.
    \label{eq:update_param}
\end{equation}
Here $\langle \cdot \rangle_{i}$ represents some mean-field parameter at the $i$th iteration, and $\langle \cdot \rangle_{i}^{\text{new}}$ its new value after performing one iteration step. The update parameter $\eta$ lies in the interval $(0,1]$.

\begin{figure}[t]
\includegraphics[width=0.8\columnwidth]{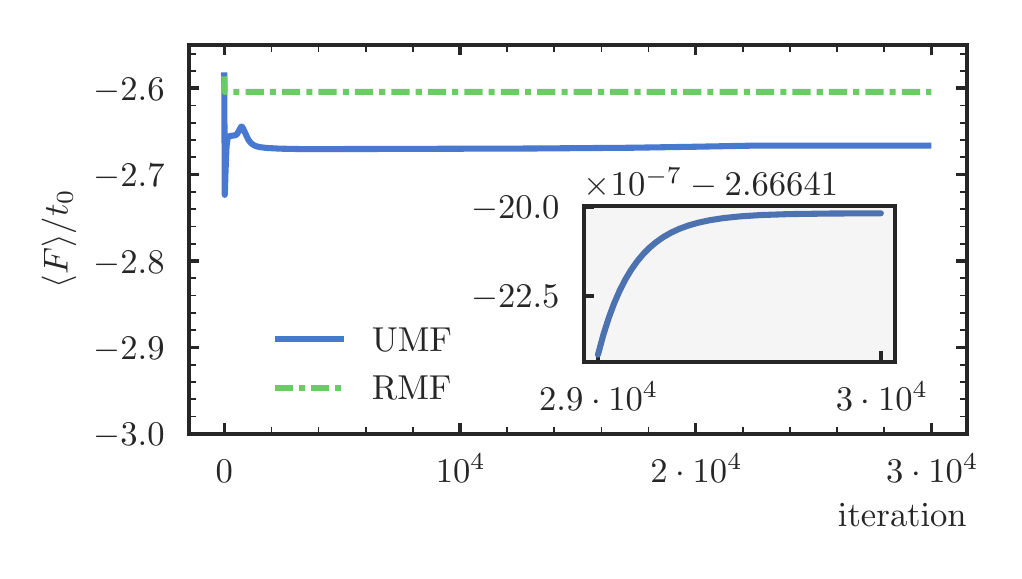}
\caption{Example of the free-energy evolution during a self-consistent loop. This figure corresponds to the residual interactions model with the same parameters as in Fig. 2 of the main text, at $T=174 \text{K}$.}
\label{fig:conv}
\end{figure}

Each unrestricted solution has been obtained after ${\sim}3\times 10^4$ iterations (see Fig.~\ref{fig:conv}), starting from noisy homogeneous distributions of $\langle u_b^2 \rangle$ and $\braket{n_{i,\sigma}}$. The fact that the mean-field parameters evolve towards non-homogeneous patterns reflects the meta-stability of the homogeneous ansatz. The update parameter $\eta$ has been initialized at $\eta=0.03$ and progressively increased until reaching the value $\eta=1$ for the last ${\sim}3\times 10^3$ iteration steps. The variation in the free energy in the last steps of the iteration algorithm is around $\Delta F \sim 10^{-8}t_0$. 
The restricted solution is obtained in the homogeneous unpolarized parameter space ($\langle u_x^2 \rangle$, $\langle u_y^2 \rangle$, $\langle Q_x\rangle$, $\langle Q_y\rangle$, $\braket{n_{i}}$). That is, only the breaking of the global rotational symmetry is allowed. In this case, one can take advantage of the spatial symmetry properties of the problem and express the quantities in Fourier space in order to reduce the computational task. The number of iterations needed to achieve convergence is much smaller for this case (${<}100$) and, for a given set of parameters, the converged energy is significantly higher than the unrestricted mean-field solutions.

Finally, notice that the convergence of the self-consistent algorithm only ensures that a metastable solution has been found. Thus, in order to choose between different solutions, one needs to compare their Free energies and chose the lowest one (e.g., in Fig.~\ref{fig:conv} the unrestricted mean-field solution has lower free energy than the restricted one).

The free energy of the electron-phonon system treated in MF + HF approximation can be written as
\begin{equation}
    F_{\text{MF}} = F_{\text{el}} + F_{\text{ph}} + C,
\end{equation}
where $F_{\text{el}}\ (F_{\text{ph}})$ is the free energy of the effective electron (phonon) Hamiltonian, and $C$ accounts for the energy shift due the MF + HF decouplings. For free fermions the free energy $F$ reads
\begin{equation}
F_{\text{el}}=\sum_i \left\{ \frac{\mu}{1+\exp(\frac{\epsilon_i-\mu}{K_BT})}-k_BT\ln \left[\exp\left(-\frac{\epsilon_i-\mu}{K_BT}\right)+1\right]\right\},
\end{equation}
where $\epsilon_i$ are the single-particle energies of $\tilde{H_{\text{el}}}$, and $\mu$ is the chemical potential. On the other hand, the free energy of the phonon of the bond $b$ with Hamiltonian $H_{\text{ph}}^b$ reads
\begin{equation}
F_{\text{ph}}^b=-K_BT\ln\left(\sum_i e^{-E_i/(K_BT)}\right),
\end{equation}
where $E_i$ are the energies of $H_{\text{ph}}^b$, and $F_{\text{ph}}=\sum_b F_{\text{ph}}^b$.
\newpage
\onecolumngrid

\section{Comparison between the mean field and the exact diagonalization}\label{app:table_MF_ED}

In Table~\ref{table:MF}, we compare the results in a 4-site cluster with periodic boundary conditions, obtained with exact diagonalization and a mean-field decoupling of the electron-phonon interaction.
\begin{table}[h!]
\caption{\label{table:MF}Comparison of the homogeneous ground-state properties using exact diagonalization (left columns) and a MF decoupling of the electron-phonon interaction (right columns) for different sizes of the local phononic basis. Here we work at zero $T$, half-filling,  $t_0=0.0083$, $t'=0$, $\nu=0.03$, $w=0.17$, $\chi_0=-0.0025$, and $U=0$, where we use atomic units (energy  $E_0=27.2\ \text{eV}$ and length $a_0=0.53\ \angstrom$).  For both methods the set of coherent states is used as a variational ansatz of the bond phonons to find the ground state around one of the minima of the quartic potential. The different parameters appearing in the table are the number of local phononic states taken into account (basis), the effective hopping of the electrons $t_{\text{b}} \equiv t_0-\nu \langle x^2 \rangle/4$, the ground-state energy ($E$), and the expected value of the local phonon operator ($\langle N_{\text{ph}} \rangle$). }
\begin{ruledtabular}
\begin{tabular}{ccccccc}
Basis& \multicolumn{2}{c}{$t_{\text{b}}/t_0$}&\multicolumn{2}{c}{$E/t_0$}&\multicolumn{2}{c}{$\langle N_{ph} \rangle$}\\
 \hline
$1$  & \hspace{.005cm} $0.742$ \hspace{.005cm} & \hspace{.005cm} $0.742$ \hspace{.005cm}&  \hspace{.005cm} $-5.795$ \hspace{.005cm}&\hspace{.005cm} $-5.795$ \hspace{.005cm} & \hspace{.005cm} $0$\hspace{.005cm}& \hspace{.005cm} $0$\hspace{.005cm} \\

$5$  & \hspace{.005cm} $0.786$ \hspace{.005cm} & \hspace{.005cm} $0.799$ \hspace{.005cm}&  \hspace{.005cm} $-5.855$ \hspace{.005cm}&\hspace{.005cm} $-5.819$ \hspace{.005cm} & \hspace{.005cm} $0.099$\hspace{.005cm}& \hspace{.005cm} $0.084$\hspace{.005cm} \\

$10$  & \hspace{.005cm} $0.792$ \hspace{.005cm} & \hspace{.005cm} $0.805$ \hspace{.005cm}&  \hspace{.005cm} $-5.855$ \hspace{.005cm}&\hspace{.005cm} $-5.819$ \hspace{.005cm} & \hspace{.005cm} $0.186$\hspace{.005cm}& \hspace{.005cm} $0.140$\hspace{.005cm} \\

$15$  & \hspace{.005cm} $0.796$ \hspace{.005cm} & \hspace{.005cm} $0.808$ \hspace{.005cm}&  \hspace{.005cm} $-5.855$ \hspace{.005cm}&\hspace{.005cm} $-5.831$ \hspace{.005cm} & \hspace{.005cm} $0.303$\hspace{.005cm}& \hspace{.005cm} $0.226$\hspace{.005cm} \\

$100$  & \hspace{.005cm} $$ \hspace{.005cm} & \hspace{.005cm} $0.819$ \hspace{.005cm}&  \hspace{.005cm} $$ \hspace{.005cm}&\hspace{.005cm} $-5.831$ \hspace{.005cm} & \hspace{.005cm} $$\hspace{.005cm}& \hspace{.005cm} $5.594$\hspace{.005cm} \\
\end{tabular}
\end{ruledtabular}
\end{table}

\twocolumngrid
\bibliographystyle{apsrev4-1}

\end{document}